\documentclass[aps,reprint, floatfix, amsfonts,amsmath,amssymb]{revtex4-1}
\usepackage{graphicx}
\usepackage{dcolumn}
\usepackage{bm}
\usepackage{amsfonts}
\usepackage{amsmath}
\usepackage{amssymb}
\usepackage{color}

\usepackage[colorlinks=true,citecolor=blue]{hyperref}
\hypersetup{colorlinks=true,citecolor=blue,linkcolor=red,urlcolor=blue}

\def\be{\begin{equation}}
\def\ee{\end{equation}}

\def\kv{{\bf k}}
\def\qv{{\bf q}}

\begin{document}

\title{Superfluidity in density imbalanced bilayers of dipolar fermions}
\author{Azadeh Mazloom}
\affiliation{Department of Physics, Institute for Advanced Studies in Basic Sciences (IASBS), Zanjan 45137-66731, Iran}
\author{Saeed H. Abedinpour}
\email{abedinpour@iasbs.ac.ir}
\affiliation{Department of Physics, Institute for Advanced Studies in Basic Sciences (IASBS), Zanjan 45137-66731, Iran}
\date{\today}

\begin{abstract}
We study the zero temperature phase diagram of an imbalanced bilayer of dipolar fermions. We consider perpendicularly aligned identical dipoles in two layers and  investigate the effect of population imbalance on the ground state phase at different layer spacings and average densities. 
The attractive part of the interlayer interaction could lead to the BEC-BCS crossover and the Fermi surface mismatch between two layers results in interesting uniform and nonuniform superfluid phases, which we have investigated here using the BCS mean-field theory together with the superfluid-mass density criterion.
The density imbalance reduces the pairing gap.  At low densities, where the system is on the BEC side of the crossover, this reduction is quite smooth while a dense system rapidly becomes normal at intermediate density polarizations. 
Stable homogeneous superfluidity is predicted to appear on the phase diagram when the dipolar length exceeds both the layer spacing and the average intralayer distance between dipoles, a regime which should be readily accessible experimentally. This homogeneous superfluid phase becomes unstable at intermediate densities and layer spacings.
We have also examined that these uniform and inhomogeneous superfluid phases survive when the effects of intralayer screenings are also incorporated in the formalism.
\end{abstract}
\maketitle

\section{introduction}\label{sec:intro}
The interplay between Bardeen-Cooper-Schrieffer (BCS) pairing and Zeeman field in superconductors, and several interesting phenomena associated with the pairing between two spin components of electrons with mismatched Fermi surfaces, has long been the subject of theoretical investigations~\cite{Sarma,Fulde,Larkin,Takada,Machida,Casalbuoni,Giannakis,Gorbar,GorbarB}.
Gapless excitations in a superconductor subjected to an external magnetic field, where bounded Cooper pairs coexist with unpaired normal electrons~\cite{Sarma}, has been predicted. This is usually referred to as Sarma, breached-pair superfluid~\cite{Forbes}, or internal gap~\cite{Caldas} phase. 
Another possibility with mismatched Fermi surfaces is the nonuniform superfluidity of Fulde, Ferrell~\cite{Fulde}, Larkin, and Ovchinnikov~\cite{Larkin} (FFLO) type. Unlike the conventional superconductors, Cooper pairs in the FFLO phase carry finite momentum and therefore the superfluid gap has a spatially oscillating behavior. 
In spite of the several theoretical proposals and over half-a-century of experimental quests, no clear evidence of these exotic phases has been reported in superconductors so far, because of the hindering of external magnetic fields by the Meissner effect~\cite{Zwierlein}, as well as the requirement for very clean systems~\cite{Takada,Aslamazov,Decroux,Matsuda}.
Nevertheless, the heavy fermion and iron based superconductors are very promising candidates for their observation~\cite{Matsuda,Burger,Zocco}. 
Alternative theoretical proposals, such as electron-hole bilayers in semiconductor heterostructures~\cite{Pieri,Subasi} and different flavors of quarks~\cite{Alford}, for the observation of these exotic superfluid phases also exist. 

In the past two decades, experimental breakthroughs in cooling and trapping dilute atomic gases~\cite{Anderson,Davis,Jaksch,DeMarco,Greiner,Bloch} brought new hope for the realization of several long-awaited condensed matter dreams. 
In neutral atomic gases, there is usually a full control over the sign and the strength of short range s-wave interaction, through the so-called Feshbach resonance~\cite{Timmermans,chin_rmp}. 
In fermionic systems, this tunability leads to the observation of BCS-Bose-Einstein condensation (BEC) crossover, between weakly bounded cooper pairs on one side and bosonic molecules of two fermionic atoms on the other side~\cite{ Regal,ZwierleinPRL,Kinast,Bourdel,Chin}.  
Exotic superfluid  phases has already been studied extensively in fermionic systems consisting of different fermions or hyperfine states of the same atom~\cite{Zwierlein,Liu,Sedrakian,Dukelsky,Castorina,Iskinprl,Iskinpra,Son}.
In two component fermionic systems with mass and/or population imbalance, the Sarma phase has been predicted to be stable deep into the BEC side of the resonance~\cite{Gubankova,Pao}, where a mixture of the BEC of bosonic molecules and normal fermionic atoms lead to the gapless superfluidity. 
In contrast, near the BCS side, the system is vulnerable of instability towards inhomogeneous phase-separated system, or a nonuniform superfluid state of FFLO type. 
The FFLO phase in ultracold Fermi gases with short range interactions has been predicted for optical lattices as well as for homogeneous systems, however the predicted parameter window in the phase space is usually thought to be very narrow~\cite{Mora,Mizushima,Cai,Toniolo}.

In the recent years there has been an immense theoretical and experimental interest in ultracold dipolar systems such as hetero-nuclear polar molecules~\cite{Ni,Deiglmayr,Takekoshi,Weinstein,DoyleN,BethlemPRL,BethlemPRL2,BethlemN,BethlemPRA,DoylePD,Meerakker, saeed_pra2012, saeed_anphys2014}, magnetic~\cite{Sukachev,McClelland,Miao,LuPRL,LuPRL2,Griesmaier,Berglund} and Rydberg~\cite{Li,Heidemann,Vogt} atoms. 
The long range and anisotropic interaction between dipoles, and its tunability through external fields, make dipolar systems very rich playgrounds for the realization of interesting phenomena such as BCS-BEC crossover~\cite{Pikowski,ZinnerPra}, density wave instabilities~\cite{Yamaguchi,Marchetti, emre_arxiv} and topological states~\cite{Levinsen,Xu}. 
Moreover, imbalanced dipolar systems may also pave the way for the observation of above mentioned exotic superfluid phases~\cite{Baarsma}.  
 
Here, we are very interested to explore such exotic phases in imbalanced dipolar systems. For this purpose, we consider identical fermionic dipoles in a bilayer geometry, where only the populations of two layers are different. Dipoles are aligned perpendicular to the layers by an external electric or magnetic field, depending on the nature of their moments. Layered geometry suppresses chemical reactions originating from the attractive part of the dipole-dipole interaction~\cite{baranov_chemrev}. On the other hand, the layer indices act as pseudo-spin degrees of freedom and the problem could be treated within the standard BCS mean field theory. We begin with the Hamiltonian of the system, keeping only the pairing order parameter between two layers, which originates from the attractive part of the interlayer interaction. Considering fixed number of dipoles in each layer, we obtain the gap function and the chemical potentials of each layer, self consistently. 
Then, we calculate the superfluid mass density, whose sign determines instabilities of the uniform superfluid state towards nonuniform superfluid phases~\cite{Iskinprl,Iskinpra}. 
In this way, we find the phase diagram of the imbalanced bilayer system in the plane of density-polarization and intralayer coupling strength, for different values of the spacing between to layers. 
We finally use the random phase approximation (RPA) to investigate the effects of many-body screening on the phase diagram.  

The rest of this paper is organized as follows. 
In Sec. \ref{sec:model}, we introduce our model, describe how we obtain the superfluid order parameter, and the mass density to examine the stability of superfluid phase. In Sec.~\ref{sec:result} we illustrate our numerical results for different physical quantities as well as the zero temperature phase diagram of an imbalanced bilayer dipolar system. We summarize and conclude in Sec.~\ref{sec:conclusion}. 
Finally, we have devoted an appendix to discuss the effects of screening on the order parameter and phase diagram.

\section{Theory and formalism}\label{sec:model}
In this section, we aim to present the basic theory and the criteria for the realization of different phases in an imbalanced bilayer system of dipolar fermions. To this end, we first introduce the model Hamiltonian of the system. Then, using the BCS mean field approximation we determine the normal or superfluid phases of the ground state. As already mentioned, an imbalanced system can host exotic phases such as Sarma and FFLO, which here we characterize them using the so-called superfluid mass density, introduced in the last part of this section. 

\subsection{Model Hamiltonian}\label{sec:model_H}
\begin{figure}
	\includegraphics[width=0.45\textwidth]{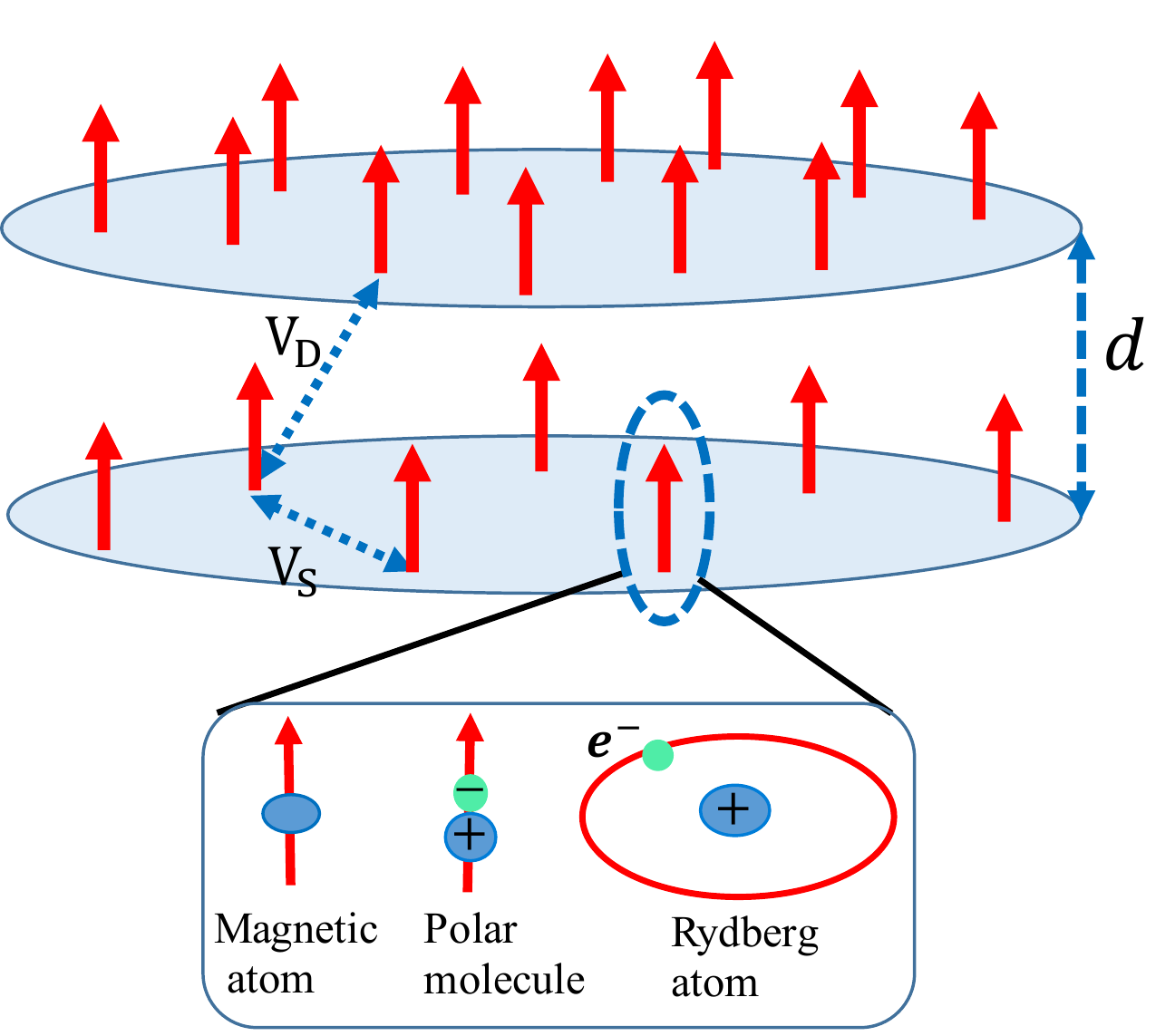}
	\caption{Schematic representation of a bilayer system of dipoles with population imbalance. The spacing between two layers $d$, intralayer interaction $V_{\rm S}$, and interlayer interaction $V_{\rm D}$ are indicated in the figure. In the lower spot, several examples of the physical systems where the dipole-dipole interaction could be relevant have been sketched.	
	\label{fig:system}}
\end{figure}

A bilayer system of dipoles which are aligned perpendicular to the planes (see, Fig.~\ref{fig:system}), is described by the following Hamiltonian 
\be\label{eq:Hamil}
\begin{split}
H&=\sum_{\kv} \xi_k^a a_\kv^\dagger a_\kv+\sum_{k} \xi_k^b b_\kv^\dagger b_\kv 
+\frac{1}{2A}\sum_\qv V_{\rm S}(q) \rho^a_\qv  \rho^a_{-\qv}  \\
&+\frac{1}{2A}\sum_\qv V_{\rm S}(q) \rho^b_\qv  \rho^b_{-\qv} 
+\frac{1}{A}\sum_\qv V_{\rm D}(q) \rho^a_\qv  \rho^b_{-\qv}  ~.
\end{split}
\ee
Here, operators $a_\kv (a_\kv^\dagger)$ and $b_\kv (b_\kv^\dagger)$ destroy (create) a dipole with momentum $\kv$ in layer a and b, respectively, $\xi^{a(b)}_k=\hbar^2k^2/(2 m_{a(b)})-\mu_{a(b)}$ is the single particle dispersion of dipoles in layer a (b), measured with respect to the corresponding chemical potential $\mu_{a(b)}$. As two layers host identical dipoles with unequal populations, we will have $m_{a(b)}=m$, but $\mu_a \neq \mu_b$. 
In Eq.~(\ref{eq:Hamil}), we have also introduced the density operators as $\rho^{a}_\qv=\sum_{\kv} a^\dagger_{\kv+\qv}a_\kv$ and $\rho^{b}_\qv=\sum_{\kv} b^\dagger_{\kv+\qv}b_\kv$. Finally, the dipolar interaction between particles belonging to the same S, and different D layers are respectively written as~\cite{Boronat}
\begin{equation}\label{eq:potentials}
V_{\rm S}(r)=\frac{C_{\rm dd}}{4\pi}\frac{1}{r^3}~,
\end{equation}
and
\begin{equation}\label{eq:potentiald}
V_{\rm D}(r)=\frac{C_{\rm dd}}{4\pi}\frac{r^2-2d^2}{(r^2+d^2)^\frac{5}{2}}~,
\end{equation}
where $C_{\rm dd}$ is the dipole-dipole interaction strength, which depends on the microscopic origin of the dipolar interaction, $r$ is the in-plane distance between two dipoles and $d$ is the distance between two layers as indicated in Fig.~\ref{fig:system}. Particles in the same layer repel each other, while the interaction between two dipoles from different layers is attractive for $r \le \sqrt{2} d$, and repulsive at larger in-plane separations. Therefore, the BCS-BEC crossover is expectable with tuning the strength of the attractive interlayer interaction~\cite{ZinnerPra,Matveeva}. The Fourier transforms of the intralayer and interlayer interactions read~\cite{Bruun}
\begin{equation}\label{eq:spoten}
V_{\rm S}(q)=\frac{C_{\rm dd}}{4}[\frac{8}{3\sqrt{2\pi}w}-2qe^{q^2w^2/2}\mathrm{erfc}(\frac{qw}{\sqrt{2}})]~, 
\end{equation}
and
\begin{equation}\label{eq:interpoten}
V_{\rm D}(q)=-\frac{C_{\rm dd}}{2}qe^{-qd}~.
\end{equation}
Here, $\mathrm{erfc}$ is the complementary error function. Note that the divergence in the Fourier transform of $V_{\rm S}(r)$ has been tackled here by introducing a short distance cut off $w$~\cite{Dukelsky}. 
In the following, we will show how the superfluid gap function of the system could be obtained within the BCS theory.

\subsection{Mean field approximation}\label{sec:MF}
In order to study the superfluidity, we utilize the BCS mean field approximation to reduce the Hamiltonian~\eqref{eq:Hamil} into a solvable single particle problem. For simplicity, we neglect the intralayer interaction and consider only the s-wave pairing between particles of different layers. Therefore, the superfluid order parameter reads
\be
\Delta_k=-\frac{1}{A}\sum_{\kv'}V_{\rm D}(\kv-\kv') \langle b_{-\kv'}a_{\kv'} \rangle~,
\ee
and the mean field Hamiltonian could be written as
\begin{equation}\label{H_MF}
H^\textrm{MF}=\sum_{\kv}\left[\xi_k^a a^\dagger_\kv a_\kv+ \xi_k^b b^\dagger_\kv b_\kv 
-\left(\Delta^\ast_k b_{-\kv} a_\kv+\textrm{h.c}\right)\right]~.
\end{equation}
Diagonalizing the above Hamiltonian with the help of the Bogoliubov transformations, gives the excitation spectrum as    
\begin{equation}\label{eq:energy}
	E_k^\pm = E_k \pm \delta \mu~,
\end{equation}
where $E_k=\sqrt{\xi_k^2+\Delta_k^2}$, $ \xi_k= \hbar^2k^2/(2m)-(\mu_a+\mu_b)/2$ and $\delta \mu=(\mu_a-\mu_b)/2$. We should note that here we are labeling the higher density layer as layer $a$ and the lower density one as layer $b$, therefore $\delta \mu\ge0$.
In contrast to the balanced system for which excitation energies have no zeros and the superfluid state is always gapped, for a system with population imbalance, the upper excitation branch $E_k^+$ has no zeros, whereas the lower excitation branch $E_k^-$, could have one or two zeros. 
Thus, with a topological phase transition, the Fermi surface of the lower excitation branch changes from a sphere into a spherical shell. Similar phase transitions has been also predicted for imbalanced systems of fermions with short range interactions~\cite{Iskinpra,Iskinprl}, and for electron-hole bilayers in semiconductor heterostructures~\cite{Subasi}.
   
The gap equation, which is obtained by minimizing the free energy of the system with respect to the order parameter, reads
\begin{equation}\label{eq:gap}
\Delta_k=-\frac{1}{2A} \sum_{\kv'} V_{\rm D}(\kv-\kv')\frac{\Delta_{k^\prime}}{E_{k^\prime}} 
\left[1-f^+_{k'}-f^-_{k'}\right]~,
\end{equation}
where $f^\pm_k=1/[1+\exp(\beta E^\pm_k)]$ is the Fermi-Dirac distribution function with $\beta=1/(k_B T)$ the inverse temperature. We consider fixed density of dipoles in each layer, so the gap equation should be complemented by the number equations for the densities of each layer
\be\label{eq:chem}
n_{a(b)}
=\frac{1}{2A} \sum_{\kv} \left[(1+\frac{\xi_k}{E_k}) f^{+(-)}_{k}+(1-\frac{\xi_k}{E_k}) (1-f^{-(+)}_{k})\right]~,
\ee
which could be solved to give the chemical potentials of two layers at fixed densities.
Now, the self consistent solutions of Eqs.~\eqref{eq:gap}-\eqref{eq:chem} allow us to obtain $\mu_a$, $\mu_b$ and $\Delta_k$ for given temperature, layer spacing, and layer densities. A finite pairing distinguishes the superfluid state from the normal one.

\subsection{The stability of the uniform superfluid state}\label{sec:super_mass}
A non zero solution for the gap function alone, is not necessarily associated with a homogeneous superfluid state as the ground state of system. 
Positivity of the superfluid mass density is a necessary condition for the stability of the uniform superfluid phase. 
For s-wave pairing, the superfluid mass density at zero temperature can be calculated from (for its more general form see, \textit{e.g.}, Refs.~\cite{Iskinpra,Iskinprl,Subasi})
\begin{equation}\label{eq:supermass}
\rho_s= m(n_a+n_b)-\frac{\hbar^2}{4\pi}\sum_{j} \frac{(k_j^-)^3}{|\frac{dE_k^-}{dk}|_{k=k_j^-}}~,
\end{equation} 
where $E_k^-$ is the lower branch of the quasi particle dispersion, as defined in Eq.~(\ref{eq:energy}), whose $j$-th zero is located at $k_j^-$. The uniform superfluid is a local minimum of the energy when the superfluid mass density is positive, whereas negative superfluid mass density guarantees the instability of the Sarma phase towards a nonuniform superfluid~\cite{Wu,He}. 
Such a nonuniform state could either refer to a phase separation between normal and paired particles, or finite momentum pairing of the FFLO type. 
In the following we will investigate different phases of the system at zero temperature, calculating the pairing gap which distinguishes the normal phase from the superfluid one, and the sign of the superfluid mass density, which characterizes the stable Sarma phase or its instability towards the FFLO phase. 

\section{Results and discussion}\label{sec:result}
Before turning to the discussion of our numerical findings, we should note that an imbalanced bilayer of perpendicular dipoles at the zero temperature could be specified by three dimensionless parameters. The density polarization $\alpha= (n_a-n_b)/(n_a+n_b)$, the dimensionless average in-plane separation between dipoles $\lambda=r_0 \sqrt{2 \pi  (n_a+n_b)}$, and the dimensionless distance between two layers $d/r_0$, where $r_0=m C_{\rm dd}/(4\pi \hbar^2)$ is the characteristic length of dipole-dipole interaction~\cite{baranov_chemrev}. 
 
Our results in this section are obtained from the self consistent solution of Eqs.~\eqref{eq:gap}-\eqref{eq:chem} at a vanishing temperature. This gives the s-wave pairing gap as a function of the wave vector $k$, together with the chemical potentials of two layers. Afterwards, we use equation~\eqref{eq:supermass} to find the superfluid mass density, which serves as the stability criterion of the uniform superfluid phase. Note that in the following we have expressed all lengths and energies in the units of $r_0$ and $\varepsilon_0=\hbar^2/(mr_0^2)$, respectively. 

\begin{figure}
	\includegraphics[width=0.45\textwidth]{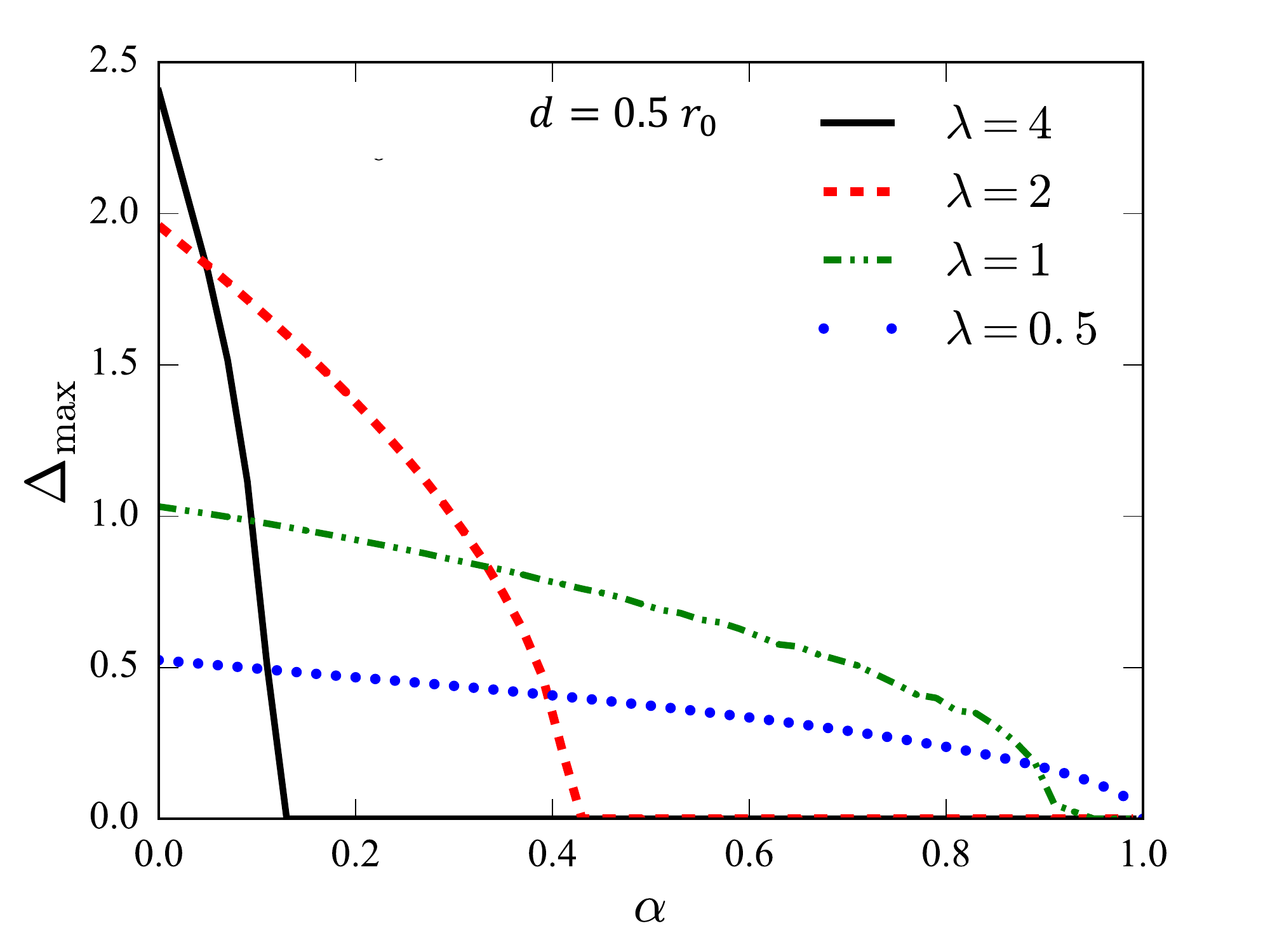}
	\caption{Maximum value of the superfluid gap function $\Delta_{\rm max}$ (in units of $\varepsilon_0$) as a function of the density polarization $\alpha$, at $d=0.5\,r_0$ and for several average particle densities. 
	\label{fig:delmax}
}
\end{figure}

\begin{figure}
\includegraphics[width=0.45\textwidth]{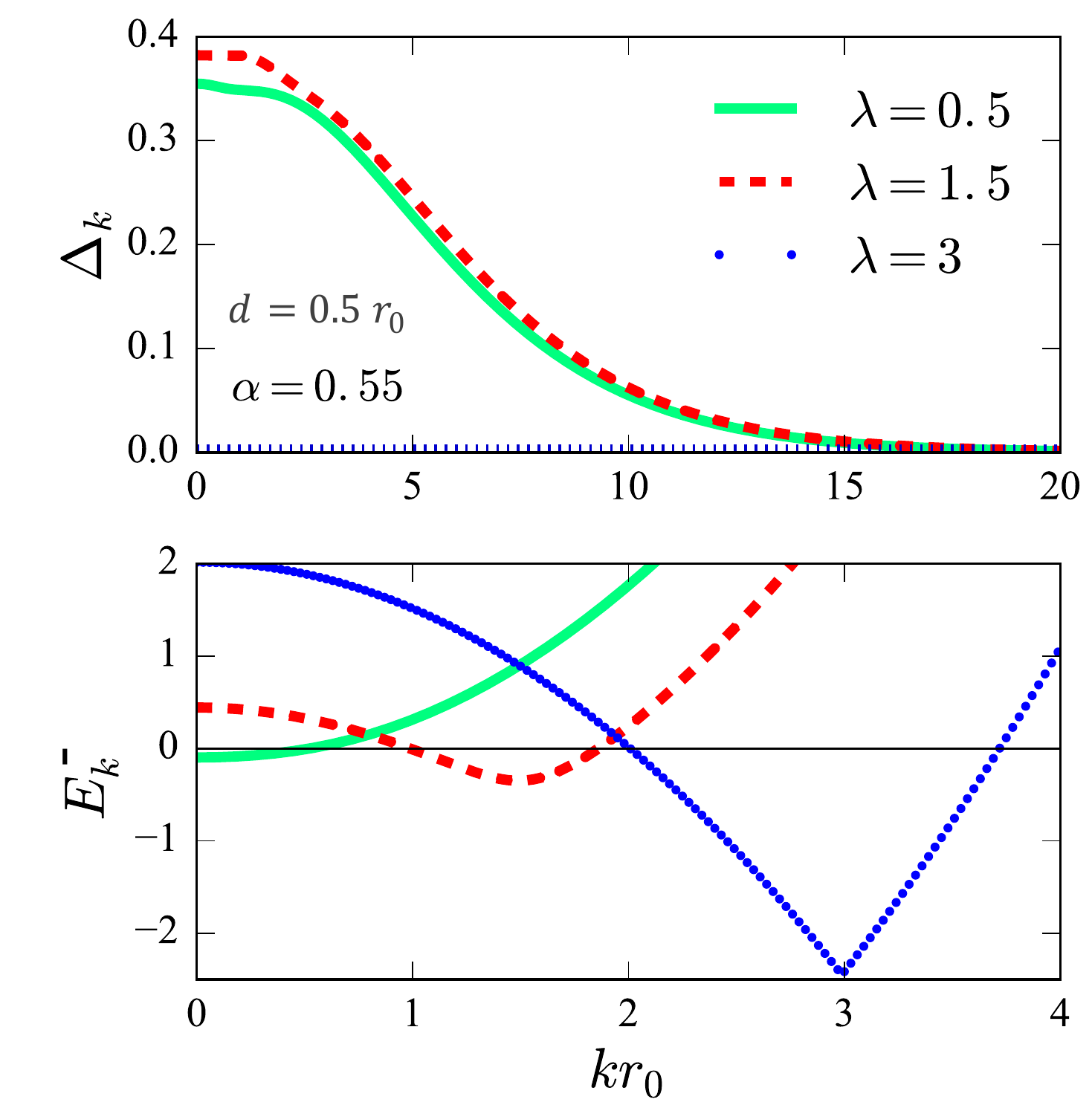}
\caption{The wave vector dependance of the superfluid gap $\Delta_k$ (top) and the lower branch of the excitation energy $E^-_k$ (bottom), in units of $\varepsilon_0$, at fixed interlayer spacing $d=0.5\, r_0$, and density polarization $\alpha=0.55$. Three different particle density parameters, $\lambda=0.5$, $1.5$, and $3$ correspond to the Sarma, FFLO, and normal phases, respectively. Note that the blue dotted line in the top panel is not easily visible, as the pairing gap is zero for the normal phase at $\lambda=3$.  
\label{fig:del_E_k}}
\end{figure}

Figure~\ref{fig:delmax} illustrates the effect of the population imbalance $\alpha$ on the pairing gap, which slowly decreases at small particle densities, while a dense system becomes normal, with a small density imbalance between two layers. 
Mismatch between two Fermi surfaces generally counteracts the pairing, but this should affect more the weak BCS-type of pairing at large densities, than the strong molecular-like binding of two dipoles at smaller densities~\cite{Mazloom_symmetric}.
In figure~\ref{fig:del_E_k} we show the wave vector dependence of the pairing gap and the lower branch of the excitation energy $E^-_k$, for fixed values of the interlayer spacing $d=0.5\, r_0$, and the density polarization $\alpha=0.55$, at different particle densities. As it is clearly noticeable, at $\lambda=0.5$ the excitation energy has one zero, meaning a spherical Fermi surface for the lower band. Whereas for $\lambda=1.5$ and $3$, the Fermi surface is a spherical shell, as the lower band of the excitation energy has two zeros. 
Zeros in the excitation spectrum together with a finite superfluid gap function, is an indication of the gapless superfluid state.
Note that the upper excitation branch of the quasi particle energy $E_k^+$, is always positive and does not carry a physical significance in our discussions here. 

\begin{figure}
    \includegraphics[width=0.45\textwidth]{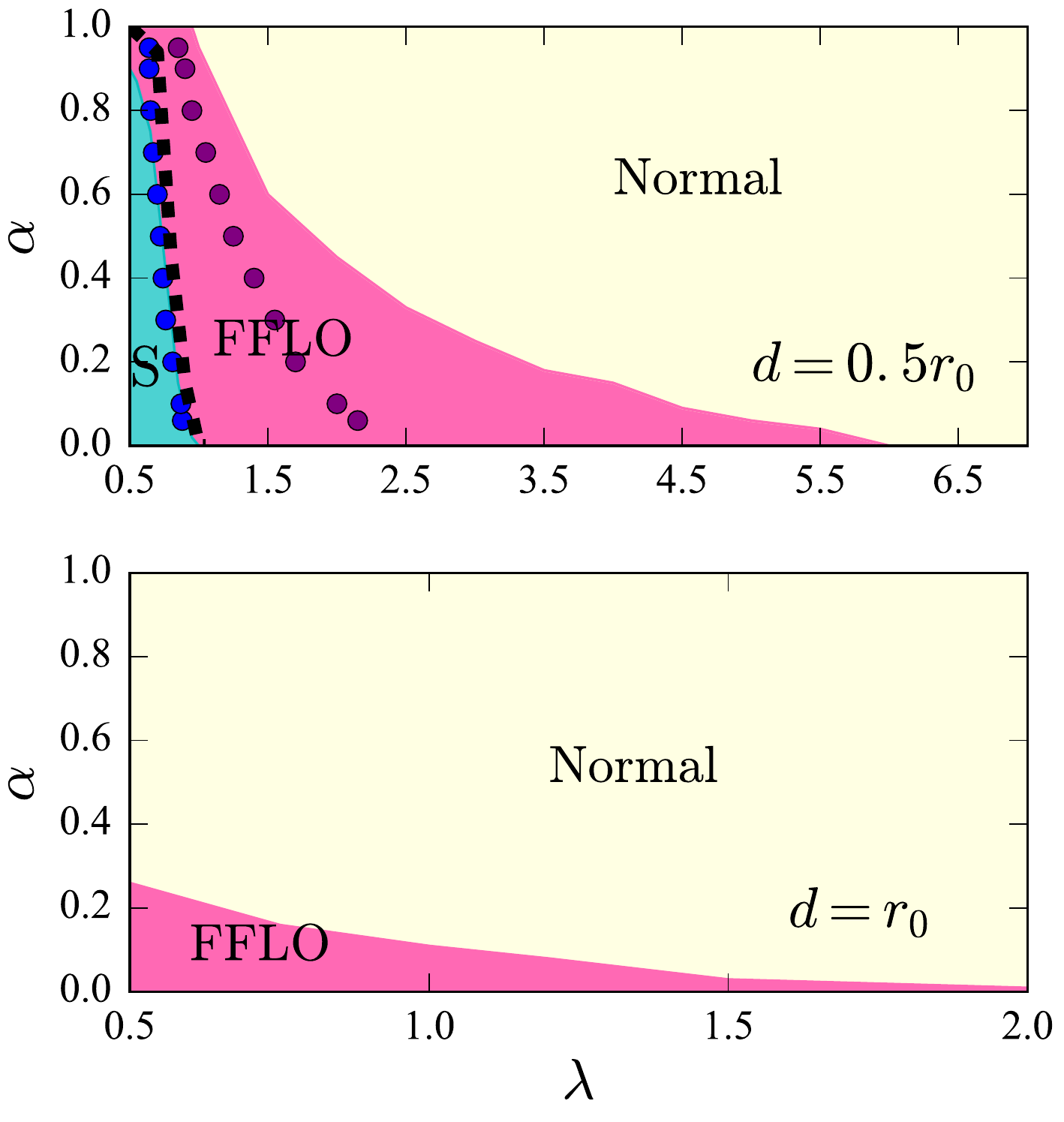}  
    \caption{The zero temperature phase diagram of an imbalanced bilayer of dipolar fermions as a function of the average density parameter $\lambda$ and the density polarization $\alpha$, obtained for interlayer distances $d=0.5 \, r_0$ (top) and $d=r_0$ (bottom). The green color shows the stable Sarma phase (S), the pink color refers to the unstable region (FFLO), and the yellow one is the normal phase, all obtained with the bare interlayer interaction. In the top panel, the blue symbols show the Sarma-FFLO phase transition boundary obtained with the RPA-screened interlayer interaction, and the red symbols define the screened phase transition boundary between FFLO and normal states. The thick dashed line in the top panel corresponds to the zero average chemical potential (\textit{i.e.}, $\mu=0$) line, such that on its left (right) side, the average chemical potential is negative (positive). 
    \label{fig:phase}
   } 
   \end{figure}

The zero temperature phase diagram of an imbalanced bilayer of dipolar fermions in the $\lambda-\alpha$ plane, and  for two different values of the layer spacing, $d=0.5 \, r_0$ and $d=r_0$, has been presented in figure~\ref{fig:phase}. 
We have observed that the superfluid order parameter is always zero for $d \gtrsim1.5 \, r_0$, irrespective of the density parameter $\lambda$, and the polarization $\alpha$. Moreover, the system is in the normal phase also at large average densities.
We found stable uniform superfluid (\textit{i.e.}, Sarma) phase only at very small layer spacings, namely at  $d\lesssim 0.63 \, r_0$. 
This is in contrast to the electron-hole bilayer for which the interlayer spacing mainly affects only the transition line between different phases, but not the number of observed phases \cite{Subasi,Pieri}. 
The black dashed line in the top panel of Fig.~\ref{fig:phase} refers to the zero average chemical potential defined as $\mu=(\mu_a+\mu_b)/2$, which separates the BEC region with negative  average chemical potential on its left side from the BCS region with positive average chemical potential on its right side. This clearly shows that the Sarma phase is stable only on the BEC side of the crossover. 

In order to verify whether these exotic superfluid phases are robust against the many-body screening due to the so-far-omitted intralayer interaction between dipoles, or not, we use the random-phase approximation~\cite{Giuliani} to find the effect of screening on the phase diagram. For this, we replace the bare interlayer potential $V_{\rm D}(k)$ in Eq.~(\ref{eq:gap}) with the screened one within the RPA  (see, the appendix for details), and repeat all the procedure to obtain the screened phase diagram. As the screening is naturally very weak at low intralayer couplings, it should not affect significantly the Sarma-FFLO phase boundary. On the other hand,  the border between FFLO and normal phases at higher densities is expected to move towards the lower densities, shrinking the FFLO phase region. This has been confirmed by our full numerical solutions, as shown by symbols in the top panel of Fig.~\ref{fig:phase}.
As it is well known~\cite{Mazloom_symmetric, neilson_prb2014}, the normal-phase RPA would generally overestimate the screening. Therefore we expect that the exact FFLO-normal phase boundary would lie somewhere between the bare and RPA results. 

\begin{figure}
    \includegraphics[width=0.45\textwidth]{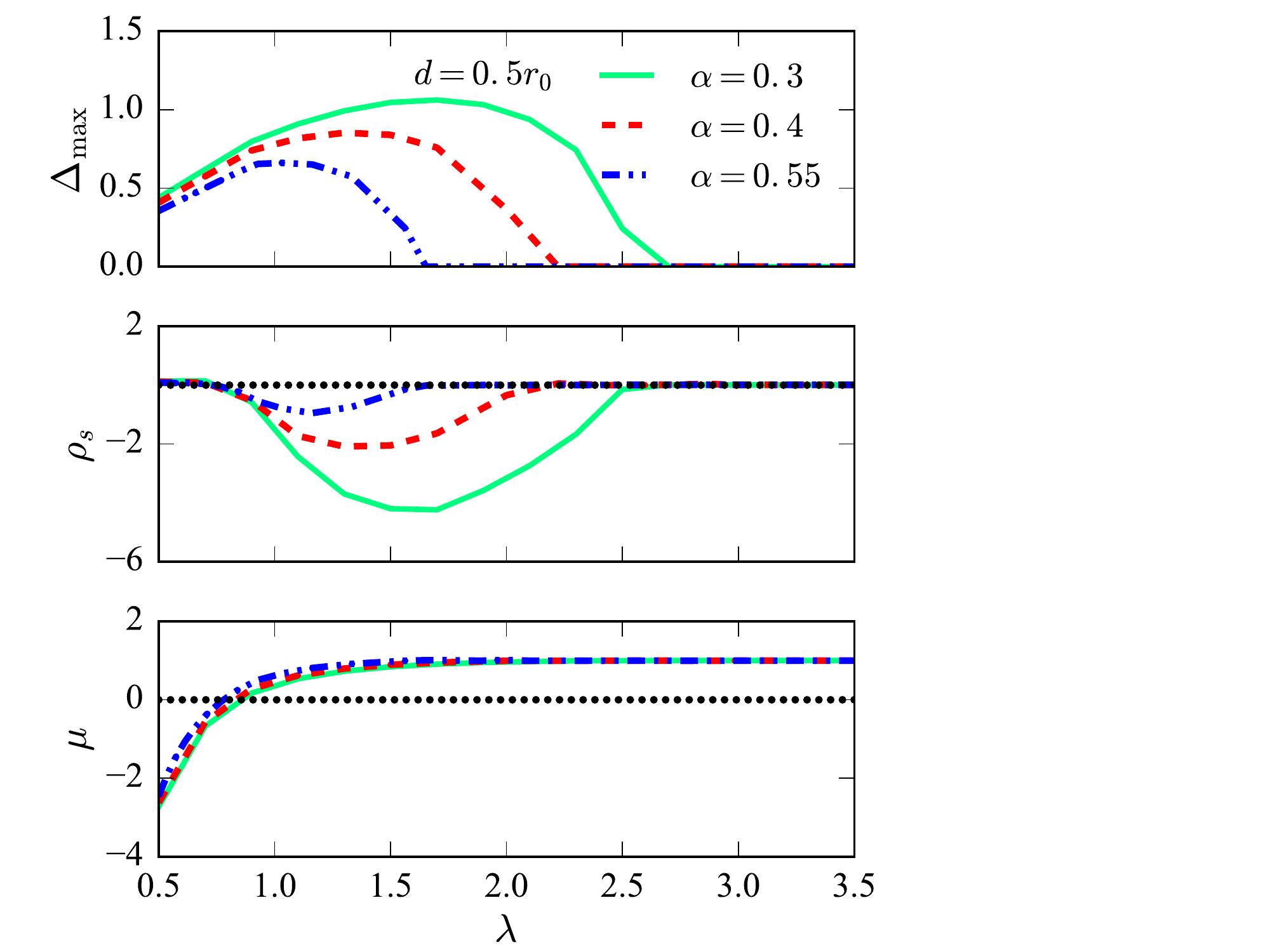}  
    \caption{Representations of, (top) the maximum value of  the superfluid gap $\Delta_{\rm max}$ (in units of $\varepsilon_0$), (middle) the superfluid mass density $\rho_s$ (in units of $m/(2\pi r_0^2)$), and (bottom) the average chemical potential $\mu$ (in units of $\lambda^2\varepsilon_0/2$), versus the average particle density parameter $\lambda$, at a fixed interlayer spacing $d=0.5\, r_0$, and for different density polarizations. 
    The superfluid-normal phase transition is characterized by $\Delta \rightarrow 0$, or equivalently $\mu \rightarrow \lambda^2\varepsilon_0/2$. 
    The sign change in the superfluid mass density indicates transition between the stable uniform, and the nonuniform superfluids states. 
    This sign change roughly co-occurs with a sign change in the average chemical potential indicating that the Sarma phase is stable on the BEC side of the crossover.
    \label{fig:del_mas_mu}
 }
\end{figure}

Finally, in Fig.~\ref{fig:del_mas_mu} we illustrate the behaviors of different system parameters across the phase transitions.  
The behavior of superfluid order parameter, superfluid mass density, and the average chemical potential has been plotted versus the average density parameter $\lambda$ for three different values of the polarization, and for $d=0.5\,r_0$. 
Vanishing of the order parameter $\Delta_{\rm max}$ specifies the normal-superfluid phase transition. 
Clearly, the average chemical potential $\mu$ in the normal region is $\lambda^2 \varepsilon_0/2$.
The stable and unstable uniform superfluid states are characterized by the sign change in the superfluid mass density $\rho_s$. Interestingly, at small density region where the Sarma phase is predicted to be stable, the average chemical potential also becomes negative which is a characteristic of an ideal Bose gas.
This suggests that the Sarma state is indeed a mixture of BEC of strongly bounded dipoles, and unpaired fermionic dipoles from the higher density layer.

\section{Summary and Conclusion\label{sec:conclusion}}
We have investigated how population imbalance in a bilayer system of dipolar fermions affects the pairing between two layers, and what exotic quantum phases appear in the zero temperature phase diagram of this system. 
We have used the BCS mean field theory to calculate the superfluid gap function, and the superfluid mass density criterion to determine the instability of the Sarma phase towards nonuniform superfluid states. 
We have shown that population imbalance reduces the superfluid gap and the superfluidity is suppressed at large population imbalances. This suppression is quite abrupt in dense systems. 
We have also obtained the zero temperature phase diagram of the imbalanced system which suggests that a bilayer system of dipolar fermions is very promising for the observation of Sarma and FFLO phases. We have finally employed the random phase approximation to examine how the area of different phases in the phase diagram are affected by the many-body screening due to the intralayer interactions. Our findings indicate that the screening pushes the FFLO-normal boundary towards the FFLO region. However, this region remains wide enough to be detectable experimentally.   
We should remind that these exotic superfluid phases require that both the distance between two layers and the average in-plane separation of particles be comparable or smaller than the dipolar length $r_0$ (see, Fig.~\ref{fig:phase}).
This regime of small interlayer spacing and low density should be readily accessible experimentally with polar molecules such as $\mathrm{NaK}$ and $\mathrm{KRb}$, whose dipolar lengths could reach few thousands of angstroms, \textit{i.e.}, comparable with the wavelength of visible light.

\acknowledgements
We thank A. L. Suba{\c s}i and L. Chomaz for helpful discussions. 
A.M. thanks IQOQI, Innsbruck, and S.H.A. acknowledges IPM, Tehran, for their hospitality during the final stages of this work.

\appendix
\section{Screened interactions within the random phase approximation}
The screened effective interaction matrix of a bilayer system within the RPA and in the static limit (\textit{i.e.}, $\omega=0$) can be written as 
\begin{equation}\label{eq:rpa}
	V^{\rm RPA}(q) = \left[1+V(q) \chi^{\rm RPA}(q)\right]V(q)~,
\end{equation}
where $V(q)$ is the $2\times 2$ bare interaction matrix
\begin{equation}
V(q)
=\begin{pmatrix}
V_{\rm S}(q) & V_{\rm D}(q)  \\
V_{\rm D}(q) & V_{\rm S}(q)
\end{pmatrix}~,
\end{equation}
with the bare intralayer and interlayer interactions being defined in Eqs.~(\ref{eq:spoten}) and ~(\ref{eq:interpoten}).
In Eq.~(\ref{eq:rpa}), $\chi^{\rm RPA}(q)$ is the matrix of static density-density response function in the random phase approxiamtion
\begin{equation}\label{rpa_matrix}
  \chi^{\rm RPA}(q)=\chi^0(q)[1-V(q)\chi^0(q)]^{-1}~.
\end{equation}
Here $\chi^0(q)$ is the matrix of noninteracting density-density response function and for a bilayer system in the normal phase ($\Delta_k=0$),  where the noninteracting interlayer response $\chi^0_{ab}(q)$ is zero, it reads
\begin{equation}
\chi^0(q)
=\begin{pmatrix}
\chi^0_a(q) & 0  \\
0 & \chi^0_{b}(q)
\end{pmatrix}~,
\end{equation}
where $\chi^0_{a(b)}(q)=-m/(2\pi \hbar^2)[1-\Theta(q-2k_{{\rm F},a(b)})\sqrt{1-(2k_{{\rm F},a(b)}/q)^2}]$  is the Stern-Lindhard function of layer $a(b)$, with the Fermi wave vector $k_{{\rm F},a(b)}=\sqrt{4\pi n_{a(b)}}$~\cite{Giuliani}.

\begin{figure}
	\includegraphics[width=0.45\textwidth]{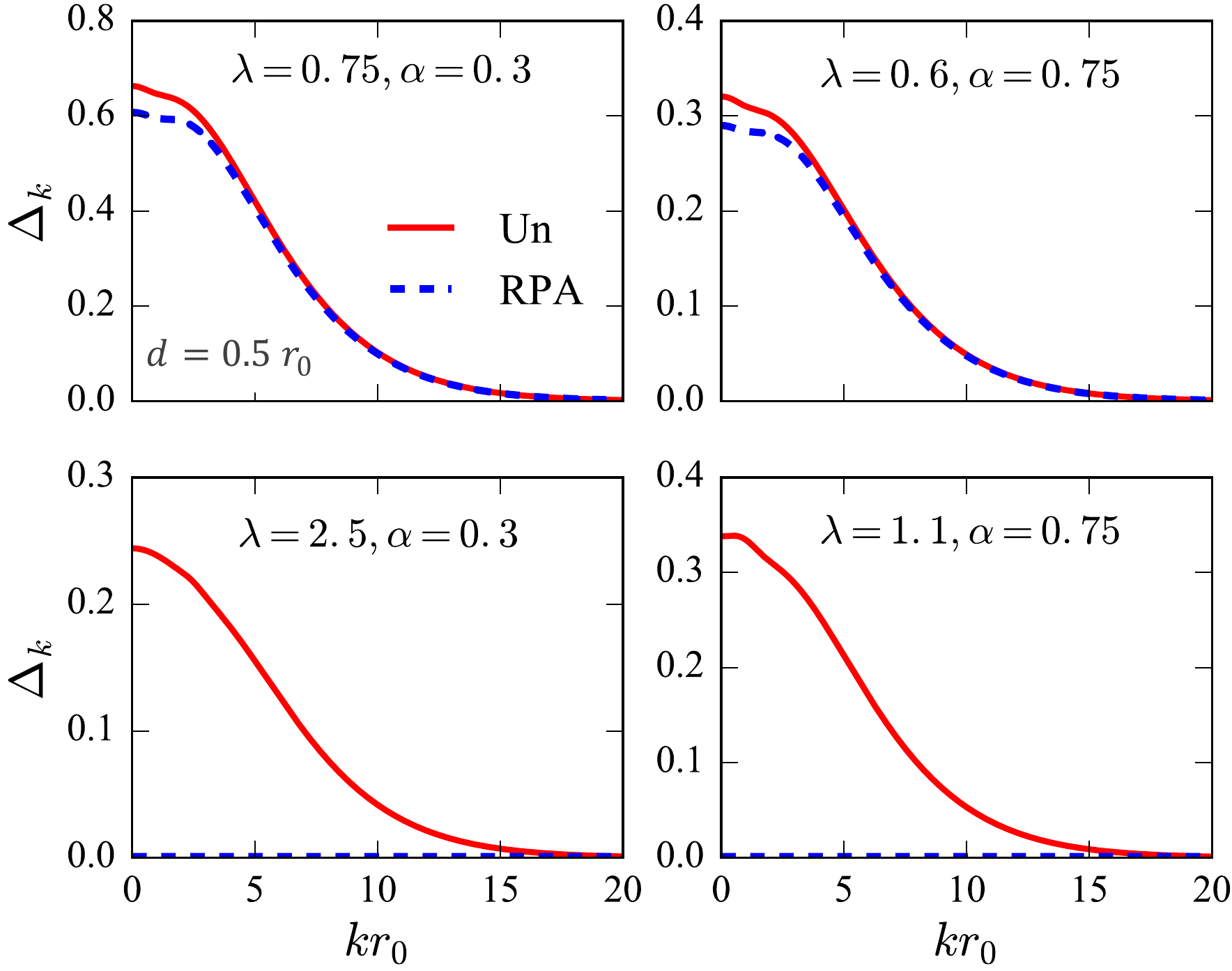}  
	\caption{The superfluid gap (in units of $\varepsilon_0$) as a function of wave vector, obtained with the bare interaction (red solid lines) and with the screened interaction within the RPA (blue dashed lines) at $d=0.5\, r_0$ and for different polarizations and particle densities.
	\label{fig:RPA_UN}}
\end{figure}

As the intralayer interaction $V_{\rm S}(q)$ depends on an artificial short-distance cutoff parameter $w$, we improve upon the RPA, with the help of the Hubbard local field factor $G_{\rm H}(q)=V_{\rm S}(\sqrt{k_{\rm F}^2+q^2})/V_{\rm S}(q)$~\cite{Giuliani}, replacing the bare intralayer interaction with
\be\label{eq:vs_hub}
\begin{split}
V^{\rm H}_{\rm S}(q)&=\left[1-G_{\rm H}(q)\right]V_{\rm S}(q)\\
&=\frac{C_{\rm dd}}{2}\left[\sqrt{k_{\rm F}^2+q^2}-q\right]~,
\end{split}
\ee
where in the second line, we have taken the $w\to 0$ limit.  
This approximation has two main benefits. First, it partially includes the effects of exchange hole, missing in the standard RPA, and second, it also removes the cutoff dependance from the model.  
Now using the screened interlayer interaction $V^{\rm RPA}_{\rm D}(q)$ instead of the bare one in the gap equation (\ref{eq:gap}), we can calculate the order parameter. In figure~\ref{fig:RPA_UN}, we have compared the pairing gaps obtained from the screened interlayer interaction with the ones of bare interaction for several values of the polarization and density. 
Evidently, at low densities where the system is deep into the BEC side of the BEC-BCS crossover, screening is negligible, while at larger densities, the RPA screening completely suppresses the superfluidity. 

\end {document}